\documentclass[12pt,iop]{emulateapj}

\usepackage{psfig}
\usepackage{apjfonts}
\usepackage{mathrsfs}

\begin{document}

\title{First evidence of fully spatially mixed first and second generations in globular clusters: 
  the case of NGC~6362}\footnotetext[1]{Based on observations
    collected with the NASA/ESA {\it HST}, obtained at the Space
    Telescope Science Institute, which is operated by AURA, Inc.,
    under NASA contract NAS5-26555. }

\author{Emanuele Dalessandro\altaffilmark{2},
Davide Massari\altaffilmark{2},
Michele Bellazzini\altaffilmark{3},
Paolo Miocchi\altaffilmark{2},
Alessio Mucciarelli\altaffilmark{2},
Maurizio Salaris\altaffilmark{4},
Santi Cassisi\altaffilmark{5},
Francesco R. Ferraro\altaffilmark{2},
Barbara Lanzoni\altaffilmark{2}
}
\affil{\altaffilmark{2} Dipartimento di Fisica e Astronomia Universit\`a di
Bologna, viale Berti Pichat 6/2, I--40127 Bologna, Italy\\   
\altaffilmark{3} INAF-Osservatorio Astronomico di Bologna, Via Ranzani 1, 40127, Bologna\\
\altaffilmark{4} Astrophysics Research Institute, Liverpool John Moores University,
IC2, Liverpool Science Park, 146 Brownlow Hill, Liverpool L3 5RF, UK\\
\altaffilmark{5} INAF - Osservatorio Astronomico di Collurania, via Mentore Maggini, 
 64100 Teramo, Italy}

\begin{abstract} 
We present the first evidence of multiple populations in the Galactic globular cluster NGC~6362.
We used optical and near-UV Hubble Space Telescope and ground based photometry,
finding that both the sub giant and red giant branches are split in two parallel sequences 
in all color magnitude diagrams where the F336W filter (or U band) is used. 
This cluster is one of the least massive globulars ($M_{\rm tot}\sim5\times10^4 M_{\odot}$) 
where multiple populations have been detected so far. 
Even more interestingly and at odds with any previous finding, we observe that the two identified 
populations share the same radial distribution all over the cluster extension. 
NGC~6362 is the first system where stars from different populations are found to be completely spatially mixed. 
Based on N-body and hydrodynamical simulations of multiple stellar generations, 
we argue that, to reproduce these findings, NGC~6362 should have lost up to the 80\% of its original mass.

\end{abstract}

\section{Introduction}
Globular Clusters (GCs) have long been considered the best example of 
{\it Single Stellar Populations}, i.e., stellar systems formed by stars with the same age and initial chemical
composition (Renzini \& Buzzoni 1986). This traditional paradigm remains still valid to
a certain extent, although a wealth of recent results showed that GCs are not as simple as
previously thought, harboring multiple stellar populations (MPs). 
Indeed star-to-star variations of light elements have been known for decades 
(see for example Cohen 1978), but this evidence was limited to a small number of stars in each cluster and 
was still compatible with a pure evolutionary effect (mixing).    
Only in the last years, intense and extended spectroscopic campaigns of large samples of stars in any
evolutionary stage, have established, 
with a high degree of confidence, that all GCs show primordial star-to-star variations
in C, N, O, Na, Mg and Al\footnote{Two exceptions to this rule has been recently found: 
Ruprecht~106 (Villanova et al. 2013) and Terzan~8 (Carretta et al. 2014), which are both very low-mass clusters.},
and a strict homogeneity in the iron abundance\footnote{
GC-like stellar systems with clearly distinct multi-modal iron distributions are 
only $\omega$ Centauri (see for example Johnson et al. 2010 and references therein) and
Terzan~5 (Ferraro et al. 2009; Origlia et al. 2013). Additional clusters with a measurable spread
in [Fe/H], confirmed with high resolution 
spectroscopy, are M~54 (Carretta et al. 2010a), M~22 (Marino et al. 2011) and M~2 (Yong et al. 2014).} 
(Gratton, Sneden \& Carretta 2004; 
Carretta et al. 2010). 
%The observed variations of light elements, indicate that a fraction of stars in GCs must ahev been
%formed out of matter processed though a high-temperature CNO cycle in a first generation of stars (FG; Carretta et al .2009 and
%references therein).

Such chemical inhomegeneities produce a variety of features in the color magnitude diagrams (CMDs)
as highlighted by high quality photometry in appropriate filters. 
For example, main sequence splittings, 
as those observed in $\omega$~Centauri (Bedin et al. 2004), NGC~2808 
(Piotto et al. 2007) and NGC~6752 (Milone et al. 2010), are thought to be originated by large spreads 
in the He abundance,
while Sub Giant Branch (SGB) splittings can be driven by differences in the C+N+O abundances 
(e.g. Milone et al. 2008; Cassisi et al. 2008). However, the most ubiquitous photometric tracer
of multiple populations is the color spread and/or the multi-modality in red giant branches (RGBs) 
with near-UV filters (NUV; see for example  
Marino et al. 2008; Lardo et al. 2011; Monelli et al. 2013). 
In these CMDs, stars enriched in C and N all lie on the red side of the RGB, while 
stars with pristine abundance of these elements lie on the blue side. This effect is driven by the 
relative strength of
deep CN and NH features in the range 3000\AA$<\lambda<$4000 \AA ~of the RGB spectra, as conclusively demonstrated 
by Sbordone et al. (2011)\footnote{However there seems to be at least one case (NGC~2419) where the effect of 
a large spread in He overwhelms that of CN and NH, inverting the color of the two populations,
(see Beccari et al. 2013, and references therein).}.
Thanks to this property, MPs have been identified in a good number of clusters so far, even by using ground-based 
photometry (see for example Monelli et al. 2013).

\begin{figure*}
\includegraphics[width=160mm]{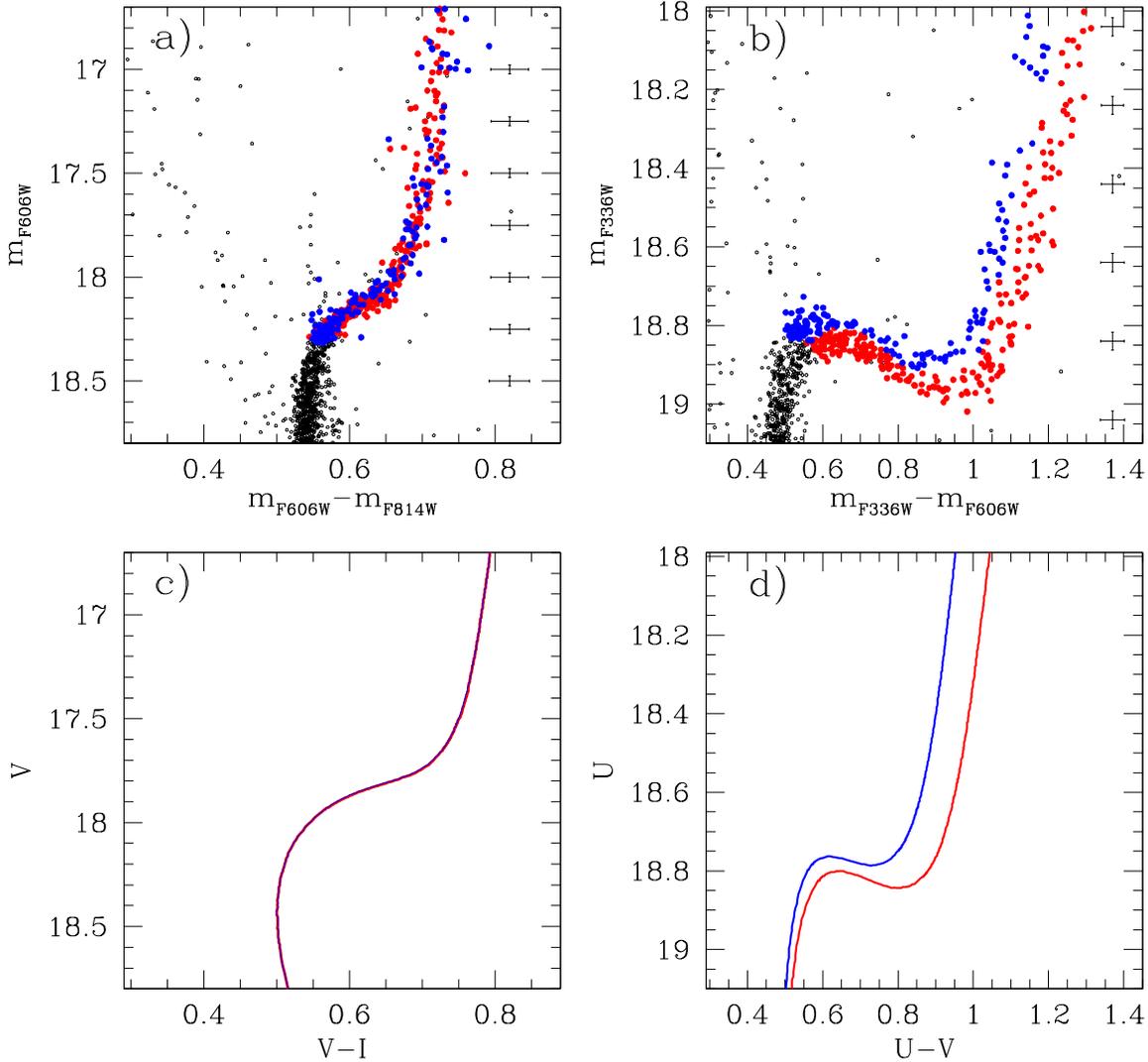}
%\vspace{3.5cm}
\caption{Zoomed view of the $(m_{F606W}-m_{F814W};m_{F606W}$; panel a) and 
$(m_{F336W}-m_{F336W};m_{F606W}$; panel b) CMDs obtained with the HST sample. Error bars as a function of
magnitude are also shown. Red and blue dots are respectively FG and SG stars. In panel c and d theoretical models 
with different light-elements mixture (Sbordone et al. 2011) and similar metallicity to NGC~6362 are shown for comparison.
The blue isochrone has a standard light-element composition, while the red one is appropriate for a Na-rich
population. }
\label{fig1}
\end{figure*}

%%framework
It is now commonly accepted that the CN-weak, Na-poor (hereafter, Na-poor) 
population is associated to a first generation (FG) 
of stars and the CN-strong, Na-rich (Na-rich) to a second generation (SG) 
formed during the first few $\sim$ 100 Myr of the cluster life, from 
intra-cluster medium polluted by the FG stars (D'Ercole et al. 2008; Carretta et al. 2009; see, however, 
Bastian et al. 2013 for a model that does 
not require multiple episodes of star formation). 
Intermediate-mass asymptotic giant
branch stars (IM-AGBs: Ventura \& D'Antona 2008), fast rotating massive stars (FRMSs;
Decressin et al. 2007), massive binary stars (De Mink et al. 2009) have been proposed as 
the most likely polluters of the primordial intra-cluster medium, enhancing Na, Al and He and depleting 
Mg and O,  while leaving the iron abundance unaffected. 
However, all formation models face
serious problems and we are still far from a full understanding of the processes responsible for the presence 
of MPs in GCs (see, e.g., Renzini, 2008). 
%In fact, in order to match the observed chemical patterns, 
%both formation scenarios require non standard initial mass-functions or GCs to have 
%been initially 10-100 times more massive than observed today (typically $M_{GC}\sim10^5M_{\odot}$).

D'Ercole et al. (2008) performed hydrodynamical and N-body simulations to explore the formation 
of MPs in the 
scenario of IM-AGBs polluters. In their analysis, AGB ejecta form a cooling flow that rapidly 
collects in the innermost regions 
of the cluster, forming a concentrated SG stellar subsystem (see also Bekki 2011). 
Their models predict also that, in the early evolutionary phases,
the cluster looses a significant fraction of its original mass.
%thus accounting for the common observational evidence of systems in which the
%SG is equally or more populous than the FG.
Indeed the early explosions of SNII lead to a strong and preferential loss of FG stars,
resulting in a cluster with a similar number of first and second generation stars (D'Ercole et al. 2008).
For comparison we refer the reader to the constraints on the mass-loss budget of GCs in dwarf galaxies  
obtained by Larsen et al. (2014 and reference therein).

After the early loss of FG stars, 
%in the resulting MP cluster 
%SG stars are 
%more centrally concentrated than the others.
the system eventually starts its long-term dynamical evolution
driven by two-body relaxation.

Understanding the dynamics and the characteristic time-scales over which MPs retain memory of their primordial
distribution is crucial to properly constrain the models and get new precious insight on GC formation and
evolution.
In all cases where the radial distribution has been analyzed, 
the Na-rich population has been invariably found 
to be more concentrated than its Na-poor counterpart
(see e.g. Bellini et al. 2009; Carretta et al. 2010a; 
Kravstov et al. 2011; Lardo et al. 2011; Johnson \& Pilachowski 2012; 
Beccari et al. 2013).

In this work we extend such kind of studies to the low-mass GC NGC~6362.
By means of a proper combination of high resolution Hubble Space Telescope (HST) and ground based optical
and NUV data, we revealed the presence of multiple sequences along the SGB and RGB 
of this cluster. Surprisingly, we have found that the two populations do not show any difference in their
relative radial distribution all over the cluster extension. 
This result represents the first evidence of fully spatially mixed multiple populations ever collected in a GC.

\begin{figure}[]
\includegraphics[width=90mm]{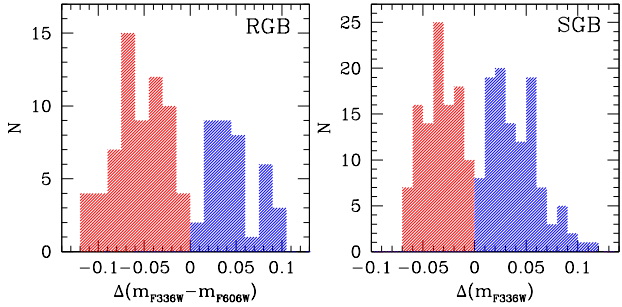}
%\vspace{3.5cm}
\caption{$\Delta(m_{F336W}-m_{F606W}$) color distribution of RGB stars (left panel) and
$\Delta(m_{F336W})$ magnitude distribution of SGB stars (right panel) 
with respect to  the adopted fiducial line (Section~3).}
\label{fig2}
\end{figure}

\section{Observations and Data Analysis}

The {\it high-resolution} HST data-base consists of images collected in
two different observing-runs. In the first one (HSTa; Prop: GO10775, PI: Sarajedini), Advanced Camera for
Survey (ACS) Wide Field Channel (WFC) images  were obtained 
with the F606W %($4\times130$ sec and $1\times 10$sec) 
and F814W filters.
%($4\times150$ sec and $1\times 10$sec).  
The second set (HSTb; Prop: GO12008, PI: Kong)
is a combination of ACS/WFC data secured with the F658N %(1 image of 750 sec and one of 766 sec) 
and F625W filters %(1 image of 140 sec and one of 145 sec), 
and Wide Field
Camera 3 (WFC3) UVIS channel images obtained through the F336W filter.
%($5\times450$ sec and $1\times 368$ sec).
%For the analysis we used flat-fielded, bias-subtracted Charge Transfer Efficiency corrected 
%images 
%(${\rm \textunderscore flc}$) for both ACS and WFC3 data.
%Pixel Area Map effects have been corrected by using the most updated PAM images available
%at the HST web-site. Geometric distortions have been corrected by using equations by Sirianni et al.
%(2005) and Bellini et al. (2011)
As done in previous works (see Dalessandro et al. 2014a) 
the data reduction has been performed independently for each data-set by using ALLSTAR
and ALLFRAME (Stetson 1994). 
%In each image a large number of bright and isolated stars have been
%selected to model the Points Spread Function (PSF). Then, we built different
%star master-lists for the two data-sets. For both  HSTa and HSTb, the master-list  was
%composed by stars properly measured in at least four (out of ten) frames. Starting from
%the master-lists and independently for each data-base, we re-performed the PSF-fitting 
%analysis on single images by using ALLFRAME (Stetson 1994).
%For each band, single frame catalogs have been combined by using DAOMATCH and DAOMASTER. 
Instrumental magnitudes were converted to the VEGAMAG photometric system by using
the zeropoints and equations reported by Sirianni et al. (2005) and listed on the HST web-site.
By combining HSTa and HSTb data we also performed a relative proper motions analysis 
(Massari et al. 2013; Dalessandro et al. 2013). The results presented in Section~3 and 4 are based 
on proper-motion decontaminated CMDs.

The {\it wide-field} data-base consists of images obtained with the Wide Field Imager (WFI)  
mounted at the MPG/ESO 2.2 m telescope (Prop: 71.D-0220(A), PI: Ortolani). 
Images were collected in three bands: $U50$\textunderscore$ESO877$ (hereafter U), 
$B123$\textunderscore$ESO878$ (B) and $V89$\textunderscore$ESO843$ (V).
%Master bias and flat-fields have been obtained by using a large number of calibration
%frames. 
%Scientific images have been corrected for bias and flat-field by using standard
%procedures and tasks contained un the Image Reduction and Analysis Facility (IRAF).
The analysis has been performed following the same approach used for the HST data.
We reported the $U$ and $V$ instrumental magnitude to the VEGAMAG photometric system by using
the stars in common with HSTa and HSTb.
The B magnitudes were reported instead to the standard Johnson photometric system 
by using photometric standards.
%the stars in common with the photometric catalog by Stetson (2000).

\begin{figure}[]
\includegraphics[width=85mm]{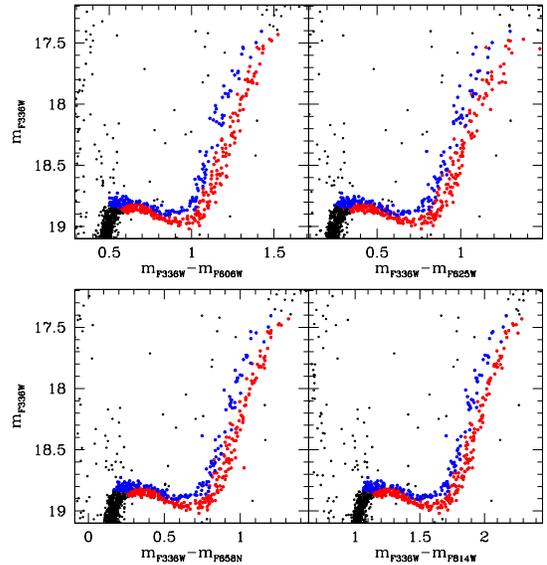}
%\vspace{3.5cm}
\caption{Different CMDs obtained combining the F336W filter with all the other bands of the HST sample. 
It is worth noticing that the blue and red populations remain clearly separated in all the CMDs.}
\label{fig3}
\end{figure}

\begin{figure*}[]
\includegraphics[width=160mm]{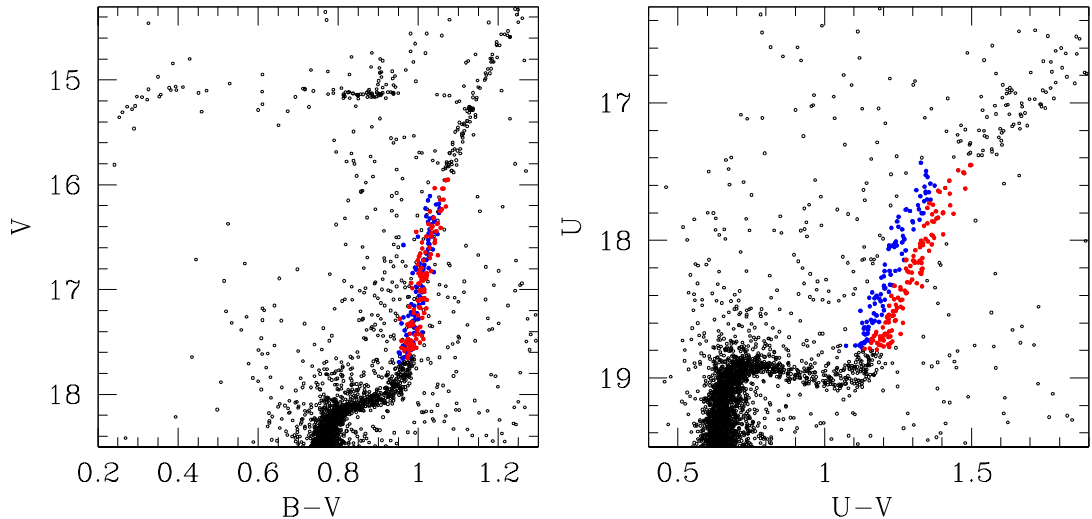}
%\vspace{3.5cm}
\caption{Optical (left panel) and NUV (right panel) CMD of the WFI sample for $r<300\arcsec$.
Red and blue dots are selected on the basis of their location in the NUV CMD, in the magnitude range $17.4\leq U \leq18.8$}
\label{fig4}
\end{figure*}

Both the {\it high-resolution} and {\it wide-field} data-sets have been put on the absolute astrometric system 
(Dalessandro et al. 2013).
The final catalogue is composed by stars detected in both HSTa and HSTb
and in the complementary WFI field of view.

Details about the data-reduction procedures, proper-motion analysis and catalogs construction will
be reported in forthcoming paper (E. Dalessandro et al. 2014b, in preparation).

\section{Multiple Populations in NGC~6362}

In Figure~1 (panel a) we show a zoom of the $(m_{F606W}-m_{F814W}, m_{F606W}$) CMD of NGC~6362. 
All the main evolutionary sequences 
(MS, SGB and RGB) are well defined and their typical 
width is fully compatible with photometric errors.
Conversely, in the $(m_{F336W}-m_{F606W}, m_{F336W}$) CMD (Figure 1, panel b) the SGB and the 
RGB broaden or clearly split. 
In particular, starting from the turnoff it is possible to distinguish 
two sequences and follow them at different magnitude levels.
We have verified that such a behavior cannot be due to differential reddening 
(Dalessandro et al. 2014b, in preparation). 
We determined a fiducial line for the observed $(m_{F336W}-m_{F606W}, m_{F336W}$) CMD.
As shown in Figure~2, the distance of the stars from this fiducial line, in the 
$(m_{F336W}-m_{F606W}$) color for the RGB and the
$(m_{F336W}$) magnitude for the SGB, is clearly bimodal. 
We used these distributions to select 
the red and blue sub-populations. 
As expected, the two populations are perfectly superimposed one to the other in the 
$(m_{F606W}-m_{F814W}, m_{F606W}$) plane (see panel a in Figure~1). 
Instead, the separation in two well distinct sub-populations remains robust in all CMDs 
where the F336W filter is used (see Figure~3).

According to combined spectro-photometric works (see for example Marino et al. 2008),
the red sequence is expected to be populated by Na-rich stars,
while the blue one is made of Na-poor stars, which should trace the primordial abundances of the clusters
(hence, the first generation).
In Figure~1 (panels c and d) we show two 12 Gyr isochrones of metallicity similar to that 
of NGC~6362 ([Fe/H]$=-0.99$; Harris 1996, 2010 edition) and with different
light-element mixtures (Sbordone et al. 2011).
Clearly the behavior observed for NGC~6362 at both the SGB and RGB levels 
in both the CMDs is qualitatively reproduced by theoretical models.
%It is interesting to note that (a) there is no sign of MS broadening or splitting in 
%our data apart from the effect of photometric errors, and (b) the CMD shows that the cluster 
%is as old as the oldest Galactic GCs 
%($t=12.5-13$Gyr) in agreement with findings 
%by Dotter et al. (2010).

%We selected red and blue stars following their separation in the 
%$(m_{F336W}-m_{F606W}, m_{F336W}$) CMD from the turnoff
%up to ($m_{F336W}=17.4$; Figure~1 c). 

In the WFI sample the same behavior emerges: in the ($B-V,V$) plane, stars define narrow sequences 
whereas in the ($U-V,U$) CMD both the SGB and RGB split or broaden more 
than expected from photometric errors (Figure~4). 
Of course
the poorer photometric accuracy and the Galactic field contamination make the distinction of the two populations 
less clear than in the {\it high-resolution} sample.
Hence, in order to study the radial distribution of the FG and SG stars, in both the HST and the WFI samples, we
considered only objects along the RGB in the magnitude interval $17.4\leq U\leq 18.8$ in the ($U-V, U$) CMD (Figure~4).

\section{Radial distribution of multiple populations}

In Figure~5 (upper panel) we show the cumulative radial distribution of 
FG and SG stars from the center to 
the tidal radius ($r_t=841\arcsec$, corresponding to $30.9 pc$ at the 
adopted distance $d=7.6 Kpc$ -- Harris 1996)\footnote{The tidal radius,
as well the other structural parameters of NGC~6362 have been obtained by fitting the surface density profile
with a single-mass King model following the approach described in Miocchi et al. (2013). 
%The best-fit is
%obtained with $r_c=67.7\arcsec$ and $c=1.09$, which give a tidal radius $r_t=841\arcsec$. These
%values are in good agreement with those found in the literature. 
Results and details obtained from this analysis will be
shown in Dalessandro et al. (2014b, in preparation).}.  
Clearly, the radial distributions of the two populations
do not show any significant difference.  According to the K-S test, the probability that the two
samples are not extracted from the same 
parent population is $P\sim50\%$, well below the standard threshold of $P=95\%$. 
The same behavior is confirmed also when the {\it high-resolution} 
and {\it wide field} samples are analyzed separately. 

We have also analyzed the radial trend of the ratio between the number of SG and FG stars 
($N_{\rm SG}/N_{\rm FG}$) as a function of the distance from the
cluster center. We divided the field of view covered by our data-base in five concentric annuli.
As shown in Figure~5 (lower panel), a flat behavior around $N_{\rm SG}/N_{\rm FG}=1.2$ 
all over the cluster extension is observed. We have carefully checked that these results are robust to meaningful 
variations in the selection of the two populations.

%Even with the high photometric quality of the data used, the separation 
%between the red and blue populations is not always clear and univocal. This is mainly due to the low number %statistics in the {\it
%high-resolution} sample and field contaminations in the {\it wide-filed} one.
%Moreover with the available set of filters we cannot perform any meaningful color-color based selection to better 
%separate them as done by Milone et al. 2013a,b. 
%We have checked however, that the results presented below are not 
%affected by small changes in the populations selection. 
 
{\it This is the first evidence ever collected of two sub-populations in a GC sharing the same
radial distribution all over the cluster extension.} \footnote{In the case of NGC~6752, Milone et al. (2013) 
found the same radial distribution of SG and FG stars in the innermost region ($r<2 r_h$). However SG stars
appear significantly more centrally concentrated than FG ones when the analysis is extended to larger 
distances (Kravtsov et al. 2011). On the other hand, Lardo et al. (2011) were unable to find difference 
in the radial distribution of the various populations in the two most distant clusters of their sample, 
because the quality of the available data were clearly insufficient for this purpose (see Beccari et al. 2013).}

\section{Discussion}

The observational evidence presented in this work demonstrates that the MPs in NGC~6362 are fully spatially mixed. 

Recently, Vesperini et al. (2013) analyzed by means of
N-body simulations, the long-term evolution of MPs in GCs as driven by two-body relaxation. 
The initial conditions of their simulations are clusters with almost the same number of FG and SG stars, 
with the latter
being more centrally concentrated than the former from their birth (D'Ercole et al. 2008).  
These systems have already experienced intra-cluster enrichment 
(D'Ercole et al. 2008) and have already lost mass (preferentially in the form of FG stars) 
as a result of the expansion triggered by SNII explosions. 
Following the dynamical evolution of all their simulations, 
the authors find that the difference in the radial distribution
of the two populations is progressively wiped out starting from
the cluster center and enclosing larger and larger portions of the cluster
as dynamical evolution proceeds and the mass lost budget, due to dynamical effects, becomes significant. 
In particular, the simulations by Vesperini et al. (2013) show that complete mixing 
is expected in advanced dynamical evolutionary stages, after $\sim 10 t_{rh}$ (the actual timescale depending on 
the initial conditions and in particular on
the initial concentration of SG stars), for clusters that lost at least $60-70\%$ of their mass 
during the long term dynamical evolution, regardless of  the assumed MP formation scenario\footnote{Note that the simulations by Vesperini et al. (2013)
have been performed assuming that all stars have the same mass. 
With a realistic mass function the
relaxation towards the populations spatial mixing would presumably be faster.}. 
To this amount of mass lost,  in the context in which MPs form through multiple star formation 
episodes (Decressin et al. 2007; 
Ventura \& D'Antona 2008; D'Ercole et al. 2008),
it should be added the expected $\sim 50\%$ of primordial mass loss during the early and somehow violent evolution ($t<1$ Gyr), mainly 
driven by SNII explosions (D'Ercole et al. 2008). Therefore we can argue 
that, in this framework, the present mass of NGC~6362 should be $\sim20\%$ of its original value.

From the collected data, we estimated the half mass radius $r_h\sim150\arcsec$ 
($\sim 5.6 pc$) of NGC~6362. 
We also derived its present relaxation time at $r_h$ to be $t_{\rm rh}\sim1.7$ Gyr corresponding
to about 1/7 of the cluster age ($t_{\rm age}=12.5-13$ Gyr; Dotter et al. 2010).     
These values are not peculiar in any respects, being fully comparable with those found in other
Galactic GCs (see for example Harris 1996, 2010 edition) still showing clear radial trends among MPs. 
$t_{rh}$ is actually slightly larger than the average for Galactic GCs. 
This would suggest that NGC~6362 had a dynamical history significantly different from 
other GGCs with similar present-day properties, leading
to a larger amount of mass loss.

By using the best-fit King model reproducing 
the density profile and the velocity dispersion data by Pryor \& Meylan (1993),
we have estimated  that the present day
total mass of NGC~6362 is $M_{\rm tot}=(5.3 \pm 1.5)\times 10^4 M_{\odot}$. 
This is about 2-3 times smaller than what we estimated,
by using the same approach, for two other small GCs, NGC~288 and M~4, for which spectroscopic and photometric
evidence of MPs have been obtained (for comparison see also the estimates by McLaughlin \& van der Marel 2005
and Sollima et al. 2012). {\it NGC~6362 might be the least massive cluster where MPs have been found so far}. 
This is an important result, which needs to be confirmed with a more accurate velocity dispersion measure, which  
is indeed the main source of uncertainties at this stage.
This finding would help to constrain 
the conditions for the onset of light-element self-enrichment in star clusters (Carretta et al. 2010b).  
The original mass of NGC~6362 should be fixed around some $2\times10^5 M_{\odot}$ 
in the MP formation scenarios proposed by D'Ercole et al. (2008) and Decressin et al. (2007),
or $\sim10^5 M_{\odot}$ in the framework suggested by de Mink et al. (2009) and Bastian et al. (2013).
These quantities are in any case smaller than the expected average primordial mass of Galactic GCs.  
In this context, it is also worth noting that the original mass of NGC~6362 is comparable, or even smaller,
than what estimated for many young clusters in nearby galaxies, where no observational evidence 
compatible with multiple star formation episodes has been revealed (see for example Cabrera-Ziri et al. 2014).

An high-eccentricity orbit leading NGC~6362 to several passages close to the MW center 
and through the Galactic disk, 
would have favored the loss of a large fraction of its original mass in addition to that lost for 
two-body relaxation.
Indeed, NGC~6362 has a quite shallow present day mass function (Paust et al. 2010), which can be indicative  
of a large amount of mass lost due to tidal effects (Webb et al. 2014). 
In addition, Allen et al. (2006) estimated a tidal destruction time of about 10 Gyr for this cluster
(and an average eccentricity $e\sim0.5$), by adopting a mass about two times larger than the one obtained
here. The use of our mass estimate would result in one of the shortest tidal destruction times 
of the author's sample. However, from our data, we are not able to find neither significant deviations 
from sphericity in the spatial distribution of stars, or breaks in the surface density profile that could 
suggest the occurrence of tidal disturbances (Dalessandro et al. 2014b, in preparation). 
Another possibility is that the two populations had similar 
distributions from the start. However this explanation would hardly fit in any of 
the present MP formation scenarios (Ventura \& D'Antona 2008; Decressin et al. 2007; de Mink et al. 2009; 
Bastian et al. 2013).

\begin{figure}[]
\includegraphics[width=80mm]{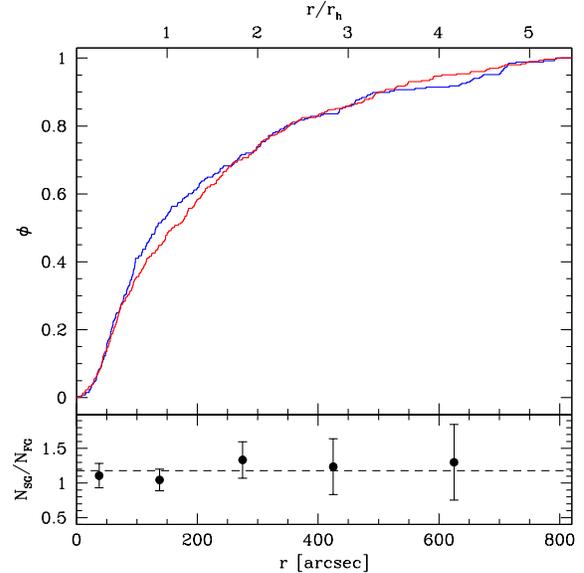}
%\vspace{3.5cm}
\caption{{\it Upper panel}: Cumulative radial distribution
of SG (red) and FG (blue) stars selected as detailed in Section~3. 
{\it Lower panel}: $N_{SG}/N_{FG}$ as a function of the distance from the cluster center. For the {\it high-resolution}
sample proper-motion selected stars have been counted, while for the {\it wide-field} sample 
a statistical decontamination from Galactic field objects has been adopted. A flat distribution around 
$N_{SG}/N_{FG}\sim1.2$ is observed.}
\label{fig5}
%\end{figure*}
\end{figure}

Indeed high-resolution spectroscopic data for a large sample of stars in this cluster is urged. In fact, 
this would allow to
fully characterize the chemical and kinematical patterns of the stellar populations of NGC~6362. 
On the other hand, a proper combination with ad-hoc N-body simulations is required to unveil the dynamical 
and star formation
history of this intriguing system.
%Similar analyses on a large number of clusters hosting MPs will shed new light 
%on their formation and evolution timescales.    
\\
\\                 
  
We warmly thank the referee Nate Bastian for his suggestions and comments 
that helped us to improve the presentation of our results. We also thank 
Enrico Vesperini for useful discussions.  
This research is part of the project COSMIC-LAB funded by the European Research Council 
(under contract ERC-2010-AdG-267675). 
SC acknowledges financial support
from PRIN-INAF 2011 "Multiple Populations in Globular Clusters: their
role in the Galaxy assembly" (PI: E. Carretta), and from PRIN MIUR 2010-2011,
project \lq{The Chemical and Dynamical Evolution of the Milky Way and Local Group Galaxies}\rq, prot. 2010LY5N2T (PI: F. Matteucci).

{}
\end{document}